\documentclass[pdflatex,sn-basic]{sn-jnl}

\usepackage[T1]{fontenc}
\usepackage{graphicx}
\usepackage{multirow}
\usepackage{amsmath,amssymb,amsfonts}
\usepackage{amsthm}
\usepackage{booktabs}
\usepackage{array}
\usepackage{xcolor}
\usepackage{textcomp}
\usepackage{url}
\usepackage{manyfoot}

\hypersetup{%
  pdftitle={When Agent Evaluation Becomes Unreliable: An Empirical Study of Oracle Failures and Configuration-Dependent Measurement in an LLM Agent Benchmark},
  pdfauthor={Junjie Chen},
  pdfkeywords={test oracle, flaky tests, measurement validity, benchmark quality, replication package, LLM agent evaluation},
  pdfsubject={Empirical software engineering; evaluation of LLM agents}
}

\begin{document}

\title[Oracle Failures and Configuration Effects in an Agent Benchmark]{%
When Agent Evaluation Becomes Unreliable: An Empirical Study of Oracle
Failures and Configuration-Dependent Measurement in an LLM Agent Benchmark}

\author*[1]{\fnm{Junjie} \sur{Chen}}\email{junjie\_chen@njust.edu.cn}
\affil*[1]{\orgdiv{School of Computer Science and Engineering}, \orgname{Nanjing University of Science and Technology}, \orgaddress{\city{Nanjing}, \country{China}}}

\abstract{%
Agent evaluation harnesses are software that run tests, generate inputs and decide outcomes with a
\emph{test oracle}. We audit defects a score table cannot show.

First, a \textbf{systematic oracle-integrity auditing approach} --- five state-integrity invariants
and three controls --- injects seven defect classes across three targets. It detects
54/57 separable target--variant combinations and 56/63 overall; the sole cross-target miss is
\texttt{G5b\_exception\_swallow}, and the un-injected implementation raises no alarm. Against a
basic-check suite built from its own components, the full suite adds 20 detections (exception
restoration, order dependence, format tolerance) at 1.7--1.8$\times$ the check cost; no check detects
over-lenient numeric matching. Upstream reports identify reward-induced state contamination
\citep{taubench_pr89,taubench_issue92}; we quantify its consequence when a harness scores after the
session. A reward-time reset leaves the reference state in place: database passes fall from 4{,}194
to 2{,}883 over 4{,}320 episodes, and success from 94.3\% to 65.5\%. The non-uniform correction
removes 24 to 34 points per agent and changes the highest point estimate.

Second, a 3{,}240-episode ablation links localisation of three documents to lower collapse in the
tested configuration, without ordering them; a second domain reverses the language effect.

Third, the method did not transfer to an independent benchmark; we name the required contract. A
replication package recomputes every reported statistic offline except the AppWorld probes, with no
model calls.}


\maketitle

\section{Introduction}\label{sec:intro}

\begin{figure}[t]
\centering
\includegraphics[width=\textwidth]{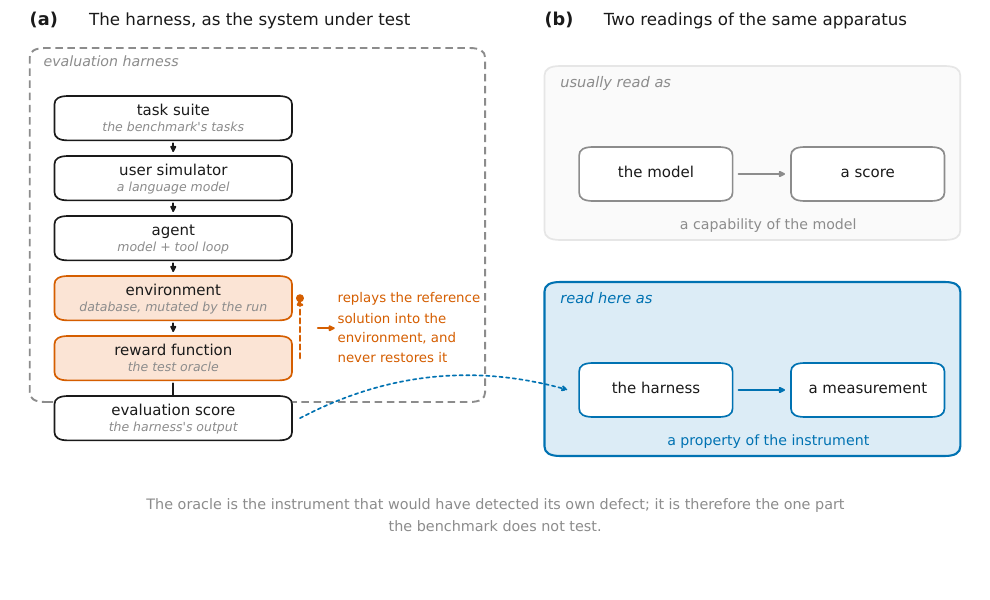}
\caption{\textbf{The evaluation harness as the system under test.}
\textbf{(a)}~An agent benchmark is a pipeline: a task suite, an input generator, the agent under
test, an environment, and a reward function that acts as the \emph{test oracle}. The oracle
re-derives the expected state by replaying the reference solution into the same environment. In the
implementation we audited it never restores what it wrote, so a harness that scores an episode
\emph{after} the session has ended reads the reference solution's state rather than the agent's
(Section~\ref{sec:res-oracle}). \textbf{(b)}~The same apparatus read two ways: the usual reading
attributes the score to the model, whereas the reading this paper takes makes the harness the
measuring instrument and the score a property of that instrument.}
\label{fig:harness}
\end{figure}

Evaluating an AI agent means running it against a benchmark and reporting the number that comes
out. That number --- a success rate --- is what a paper states, what a leaderboard ranks, and,
increasingly, what a deployment decision rests on. It is not produced by the agent alone. It is
produced by an \emph{evaluation harness}: a suite of test cases (the benchmark's tasks), a
generator of test inputs (increasingly a language-model user simulator), and a \emph{test oracle}
--- the benchmark's reward function --- that decides whether each run passed. The harness is
software, and the score is one of its outputs (Figure~\ref{fig:harness}).

Software engineering has spent decades on the failure modes of exactly this arrangement.
Deciding expected from actual is the \emph{oracle problem}, and it is where test suites fail most
quietly \citep{barr2015oracle}. A test that passes and fails on unchanged code is
\emph{flaky} \citep{luo2014flaky}. A measurement that moves when the apparatus changes,
while the property being measured does not, is a validity failure rather than a capability
difference \citep{kitchenham2002guidelines}. Fault injection, in the lineage of mutation
testing \citep{jia2011mutation}, is the standard way to ask whether a test suite would
notice a defect at all. These are settled methods with an unsettled application area.

The evaluation practice we study has adopted the software without adopting the methods. In
particular, a benchmark's reward function is treated as \emph{definitional}: it defines what success
means, so it is not itself tested. Two consequences follow. First, an error inside the oracle is
invisible by construction, because the oracle is the instrument that would have detected it, and
because the returned value can be correct --- the defect can be in what reading the oracle
\emph{leaves behind}. Second, because the harness is also how a configuration is expressed (which
language the simulator speaks, what the agent's policy and tool descriptions say), a change in the
apparatus is read as a change in the agent. Neither consequence is hypothetical: this paper
documents one instance of each through an audit of one widely used agent benchmark.

We therefore treat the evaluation harness as a system under test and audit it. We organise the work
as three questions, of which only the first is about the evaluation system itself; the other two are
analyses of the measurement process, and we state them at that strength.

\textbf{RQ1 (primary).} How reliable is the evaluation oracle --- would it notice a defect in itself,
and does it have one? The object of study is the evaluation system: the correctness and consistency
of its verdicts and the fidelity of the labels it produces. The evidence is a test suite aimed at
the oracle and a fault-injection battery aimed at the harness.

\textbf{RQ2 and RQ3 (secondary).} How do harness configurations influence observed evaluation
outcomes, and how sensitive are the conclusions to factors that vary during execution? These are
analyses of the \emph{measurement process}, run on the same audited pipeline. Their object is the
harness rather than the model: they characterise how design choices in the harness --- which
documents are localised, which task
domain is used --- expose, amplify or obscure variability that is present while a benchmark runs,
and where the resulting conclusions stop holding.

We hold this distinction throughout: \emph{evaluation-system reliability} is not \emph{intrinsic
model behaviour}. RQ1 provides direct evidence about the first. The reward side effect we study
was already reported upstream \citep{taubench_pr89,taubench_issue92}; our contribution is to measure
how it produces wrong labels and changes reported scores when an external harness performs a
post-session scoring read. RQ2 and RQ3 provide evidence about the conditions under which a
measurement outcome moves, and are reported as observations and boundaries rather than as findings
about models. The three questions carry correspondingly different evidential weight.

\paragraph{Contributions.}
\begin{enumerate}
  \item \textbf{A systematic oracle-integrity auditing approach, with sensitivity and
        behaviour under the un-injected condition measured on the implementations studied.}
        Eight checks in two families --- five state-integrity invariants and three
        known-answer controls --- plus a fault-injection battery of seven defect classes and a
        ground-truth-controlled synthetic domain. Validated across three targets; the suite
        detects 54 of the 57 separable variant--target combinations, the un-injected implementation
        raises no alarm, and removing any one of the four load-bearing detectors leaves some detected
        combination undetected (Section~\ref{sec:oracle-kit}). The three misses are the same variant on
        each target, named
        rather than summarised away (Section~\ref{sec:res-oracle}). A pre-registered ablation against
        a basic-check subset attributes 20 additional target--variant detections to INV-3, INV-4 and
        CTL-3 at about 1.7--1.8$\times$ the check cost (Section~\ref{sec:p0-incremental}).
  \item \textbf{A minimal reproduction of an upstream state-contamination side effect, and a
        replay-based correction of the labels it corrupted.} Public upstream
        reports identify reward-induced contamination of environment data and recorded actions
        \citep{taubench_pr89,taubench_issue92}. We trace the database side effect into incorrect
        post-session scoring: the benchmark's \texttt{Env.calculate\_reward()} resets the
        environment database to build a ground-truth hash and never restores it. The first returned
        reward is unaffected by this particular state-contamination mechanism, because the agent's
        database hash is captured before the reset. The benchmark's own agent consumes that value
        without reading the environment again; the wrong label arises because our harness scores the
        episode \emph{separately}, after the session has ended, and therefore re-reads an environment
        left holding the reference solution. This integration failure is scoped to that second read,
        not to the upstream runner's internal score. On 4{,}320 episodes the archived labels record
        4{,}194 database passes; replaying the recorded tool calls against a clean database gives
        2{,}883, and success falls from 94.3\% to 65.5\% (Section~\ref{sec:res-oracle}). The change
        is not a uniform offset: it differs across agents by enough to reorder the top of this
        matrix, so the defect distorts comparisons between systems and not only the absolute score
        (Section~\ref{sec:res-oracle}).
  \item \textbf{A configuration-dependent reliability effect, localised to its components.}
        Under the state-corrected scoring procedure, changing the simulator's language yields no
        resolved difference in aggregate success, but it changes by a factor of two to ten the rate
        at which one agent's trajectory degenerates. Localising the agent's own world --- its policy, its
        tool descriptions and its error strings --- is the larger lever, and a per-component
        ablation over 3{,}240 episodes associates each of the three documents, localised alone,
        with a lower collapse rate in the tested configuration.
        The three effects do not combine additively: localising the tool descriptions alone is measured
        $1.11$\,pp above localising all three. We report that ordering as an \emph{observation}
        rather than a finding --- the arms come from different run periods, the contrast is the most
        exposed of the ones we report, and our own power analysis puts a contrast of that size at
        17\% power in this design (Section~\ref{sec:res-config}). The sign of the effect does not
        survive a change of task domain, which bounds it to the configuration in which it was
        measured (Section~\ref{sec:res-config}).
  \item \textbf{A replication package and the portability boundary of the method.} We release the
        auditing suite, the seven-class fault-injection catalogue, the pre-registration and
        deviation logs, and the frozen per-episode records behind every number in this paper,
        together with offline entry points that recompute the reported statistics with no network
        access, no API keys and no model calls (Section~\ref{sec:replication}). We also report, as
        a negative result, the second independently implemented benchmark on which the method did
        \emph{not} transfer, and identify the framework contract that blocks it
        (Section~\ref{sec:threats}).
\end{enumerate}

\paragraph{Scope.} The audit covers one agent benchmark, $\tau$-bench \citep{tau_bench_repo},
with a second task domain, a synthetic domain, and one further, independently implemented
benchmark on which it did not transfer. Its claim is a \emph{systematic auditing approach
evaluated on the implementations studied}, with the measured portability boundary reported as a
result. Section~\ref{sec:threats} states the limits of that scope.

\section{Background and Related Work}\label{sec:related}

\subsection{Test oracles, mutation testing and the oracle problem}\label{sec:rw-oracle}
Deciding expected from actual is the oracle problem, a foundational difficulty of software
testing \citep{barr2015oracle}. The trouble is not that oracles are absent but that
they are \emph{unexamined}: an oracle is exercised by every test run and tested by none, so its
own faults are silent. Mutation testing grew out of this observation
\citep{jia2011mutation} --- inject a fault, then ask whether the suite notices --- and
remains the standard technique for measuring what a test suite can actually detect. We take that
technique and point it at the oracle itself rather than at the program under test.

Two properties of an oracle make our variant distinct. First, it can be \emph{stateful}: an oracle
that re-derives its expectation by replaying a reference solution against a live database can
return a correct value and still leave the system in a state no other reader expects. Second, its
contract with the harness is usually implicit --- who reads the environment, and when, is not
specified anywhere --- so a second reader can be wrong through no fault of the oracle's own
runner. Our first contribution is a state-integrity formulation of the oracle problem for
machine-learning evaluation harnesses, together with the fault-injection battery that measures it.

\subsection{Flaky tests, measurement validity and configuration sensitivity}\label{sec:rw-flaky}
A flaky test passes and fails on the same code \citep{luo2014flaky}. In empirical
software engineering the same worry appears as measurement validity: a reading that responds to
the apparatus rather than to the property under study is not a weaker measurement of that
property, it is a measurement of something else \citep{kitchenham2002guidelines}. Metamorphic
testing approaches the difficulty from a third direction, checking relations that must hold
between related inputs when a direct oracle is unavailable \citep{chen2018metamorphic}.

Our second contribution sits between these three, and it needs one distinction that the flakiness
literature does not supply. Changing the language of the apparatus changes the input the agent
receives: the simulated user now writes Chinese, or the policy, the tool descriptions and the
error strings are read in Chinese. An agent may legitimately behave differently, so a difference
in outcome is read as configuration dependence rather than as evidence that the oracle is
defective. The validity question arises one step later, when such a difference is \emph{attributed} to the
system under test --- a capability claim, a leaderboard position --- while the measurement
contract holds the apparatus fixed. Our secondary experiments are therefore reported as
configuration dependence of the observed outcome and bounded to the configuration in which they
were measured; what the component ablation manipulates is the set of documents that name the same
objects, tools and error conditions in another language. In agent evaluation the analogous
instability also has a level that CI flakiness does not: the \emph{trajectory}, where a model's
multi-step behaviour degenerates non-deterministically rather than the same test returning two
verdicts.

\subsection{Testing and evaluating LLM-based systems}\label{sec:rw-llmse}
A separate strand treats a language model as the system under test, asking how to specify,
exercise and judge a system whose outputs are not deterministic \citep{hou2024llm4se},
\citep{zhang2022mltesting}. We take that framing and apply it one level down. The system
under test in this paper is not the model but the \emph{evaluation harness}: the model is the
input generator and the agent is the system the harness measures. The distinction matters because
the two are routinely conflated --- a benchmark's reward function is treated as definitional
rather than as software with failure modes --- and because a fault in the harness is invisible in
precisely the way a fault in an oracle is.

\subsection{Agent benchmarks and the reliability of their measurements}\label{sec:rw-bench}
Benchmarks are themselves software artefacts whose validity and reproducibility must be argued
rather than assumed \citep{recht2019imagenet}, \citep{kistowski2015benchmark}. In agent
evaluation the argument has mostly proceeded by measuring the measurement's noise rather than by
auditing its apparatus: model-as-judge scores vary with the judge and the prompt
\citep{judgels2026}, \citep{agentasjudge2026}; rankings shift under resampling or a change of
reported statistic \citep{difrankings2026}, \citep{rankcs2026}, \citep{onrandomness2026};
aggregate success hides what actually happened in a run \citep{beyondpass1_2026}; benchmarks can
be answered without solving the task, or fail to discriminate between systems that differ
\citep{kehkashan2026benchmarks}; and the apparatus itself revisits
established benchmarks \citep{agenteval_survey2026}, \citep{tau_bench_repo}, \citep{barres2026tau2}.
We share that concern and change its object: instead of asking whether the score is noisy, we ask
whether the instrument that produces it is \emph{correct}, using tests aimed at the instrument.

Oracle state integrity also has public precedent in the benchmark's own development. Upstream
pull request \#89 proposes opt-in fixes for strict output matching and preservation of environment
data during reward calculation; both switches are disabled by default, the patch restores data but
not the action history, and the restoration is outside exception protection
\citep{taubench_pr89}. Upstream issue \#92 documents the appended gold actions and reports an
audit of the public HAL trajectory archive: 1,523 of 2,334 airline records (65.3\%) have a gold-action
suffix, and 58 records show a complete gold tool sequence despite no agent tool call
\citep{taubench_issue92}. These reports establish prior awareness of the side effects. The evidence
here concerns their downstream integration: whether a separate post-session scorer produces wrong
labels, by how much, and with what effect on comparisons between agents.

State integrity and \emph{replay fidelity} are related but distinct requirements. The latter asks
whether re-execution reproduces the observations needed for grading. $\tau^2$-bench avoids live
state contamination by replaying both the recorded trajectory and the gold actions in freshly
constructed environments \citep{barres2026tau2},
yet reported failures show that time-dependent tool returns can change replayed messages
\citep{tau2_issue387}, and that tools committing state before versus after reporting an error can
make strict replay fail or lax replay credit a state the agent did not produce
\citep{tau2_issue608}. Our replay checks both the terminal database state and the recorded tool
observations, which addresses the second requirement for the episodes analysed here
(Section~\ref{sec:res-oracle}).

Three recent audits are close enough to require an explicit division of labour, because they
audit different objects and their evidence does not substitute for one another
(Table~\ref{tab:audits}). \citet{gazhou2026bounds} take completed run records and ask whether the
stored evidence supports the score a benchmark reports. A locked checklist assigns each record
\emph{Evidence Pass}, \emph{Evidence Fail} or \emph{Unknown}, and the score is reported as the
evidence-supported bound $[P/N,(P+U)/N]$ rather than as a point. Across five public benchmarks,
including AppWorld and $\tau^3$-bench retail, they use a locked checklist; on $\tau^3$-bench retail
they audit 300 completed records, of which one has missing post-state evidence and 24 show a
disagreement between the reward and the recorded actions. \citet{dobenchtmark2026} audit a
different contract: whether a tool's documented interface matches the state transitions its
implementation actually performs. Our object is the oracle itself and its integration with a
harness: whether the function that returns the verdict is state-clean, correct on known answers,
and detectable when it is defective.

The three are complementary, and the division matters for what each can conclude. Evidence bounds
quantify how much of a reported score the stored evidence supports, and they can flag a record
whose reward disagrees with its actions; they do not test whether the oracle would notice a defect
in itself, and they inject none. Contract audits test tools, not the reward function. Oracle-integrity
testing supplies what neither does --- known-answer controls, state invariants and a
fault-injection battery --- and measures the detection it achieves on the variants it constructs.
A benchmark can fail all three audits independently, and passing one says nothing about the other
two.

\begin{table}[t]
\caption{Three audits of agent evaluation, by object. The rows are complementary: a record-level
evidence audit, a tool-contract audit and an oracle-integrity audit answer different questions
about the same benchmark, and none of them subsumes another.}
\label{tab:audits}
\setlength{\tabcolsep}{5pt}
\begin{tabular}{@{}>{\raggedright\arraybackslash}p{0.18\textwidth}>{\raggedright\arraybackslash}p{0.24\textwidth}>{\raggedright\arraybackslash}p{0.25\textwidth}>{\raggedright\arraybackslash}p{0.24\textwidth}@{}}
\toprule
Study & Object audited & Evidence produced & Not tested \\
\midrule
Gao \& Zhou (2026) & completed run records against the reported score & three-state evidence
label and score bounds & the oracle's state behaviour; detection under injected defects \\
Bellibatlu et al. (2026) & documented tool contract against the implementation & contract
violations, per tool & the reward function and its integration with the harness \\
This paper & the oracle and its integration with the harness & state invariants, known-answer
controls, a 21-variant battery over three targets, and a replay-based correction & whether the
stored evidence behind a score is sufficient \\
\bottomrule
\end{tabular}
\end{table}

\subsection{User simulators as the measuring instrument}\label{sec:rw-sim}
Because the user's turns are generated rather than fixed, a user simulator is not a fixture of the
benchmark so much as part of the measuring apparatus, and its own behaviour affects the reading
\citep{seshadri2026lost}, \citep{sim2real2026}, \citep{ni2026usersimsurvey}. Simulators are
designed, validated and compared as components \citep{dou2025simulatorarena},
\citep{userproxybench2026}, \citep{knowsim2026}, \citep{verifiableusersim2026},
\citep{reliablesim2024}, \citep{usersim_workshop2026}, and increasingly built on real
interaction data \citep{sotopia2023}. A related line shows that the simulator's presentation ---
persona, dialect, and the demographic cues it discloses --- moves measured outcomes without
moving model capability \citep{taday2026persona}, \citep{dialectvsdemog2026},
\citep{demcues2026}, \citep{disclosuregated2026}, and that the \emph{language} of the apparatus is a
configuration of this kind: cross-lingual evaluations are sensitive to translation, locale and
prompt language \citep{maps2026}, \citep{lostinexecution2026}, \citep{babelarena2026},
\citep{polyworkbench2026}, \citep{englishforcing2026}. That literature
measures the \emph{agent} across languages; what we add is the complementary reading of the same
manipulation as a property of the apparatus, localised to its components and checked for whether
it moves success rate or only the reliability of the run.

\subsection{Replication and artefact evaluation}\label{sec:rw-replication}
Studies of the kind this paper reports are judged as much on whether a reader can rerun them as on
what they found \citep{winter2022artifacts}. Two conventions from that practice
shape what we release. First, a replication package should carry the \emph{frozen inputs}, not a
recipe for regenerating them, because a recipe rerun months later reproduces a different system.
Second, a package should distinguish what it can check from what it cannot: ours reports a
per-source verification status and counts only sources whose file was found \emph{and} whose hash
matched as verified, rather than treating an absent archive as a pass. Both choices are recorded
in the deviation log rather than presented as the original plan.

\section{Methodology}\label{sec:method}

\subsection{The evaluation harness as a system under test}\label{sec:sut}

The object of this study is not a model but the software that evaluates one, and the mapping to
testing vocabulary is direct. The benchmark supplies a suite of \emph{test cases}: tasks that
specify a user's goal and, for each, a reference sequence of tool calls that satisfies it. The
\emph{test input} is generated at run time by a language model playing the user. The
\emph{system under test} is a tool-calling agent --- a model plus a fixed control loop --- and the
\emph{test oracle} is the benchmark's reward function, which decides pass or fail by comparing the
terminal state of a database against the state the reference actions would have produced, and by
requiring that certain output strings appear in what the agent said. The \emph{assertions} are
therefore two: a state comparison that is language-neutral, and a substring match on the final reply
that is not.

Two properties of this arrangement matter for what follows. The oracle is a function of mutable
state: to construct its expectation it must execute the reference actions, which means it must
touch the same database it is grading. And the grading is a two-component conjunction, which lets
us separate the state component from the textual one when the two disagree.

\subsection{Oracle-integrity testing}\label{sec:oracle-kit}

\begin{figure}[t]
\centering
\includegraphics[width=\textwidth]{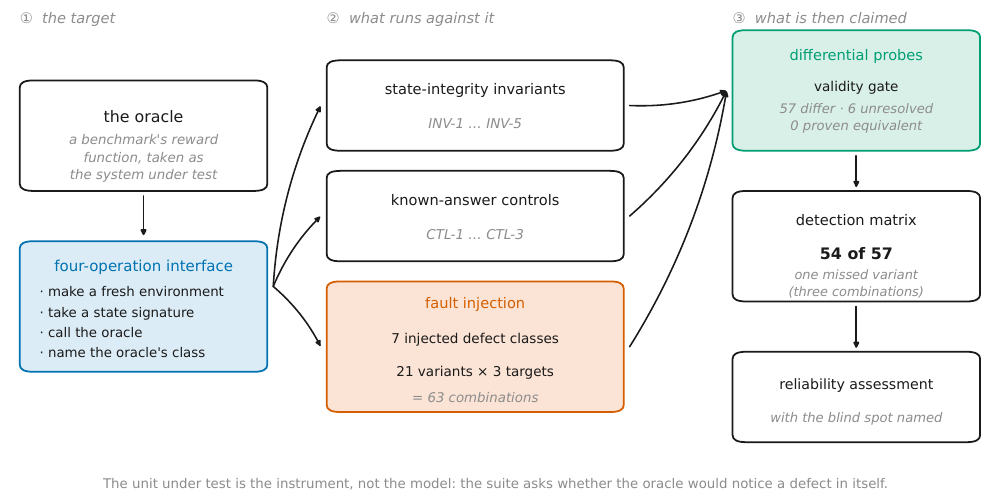}
\caption{\textbf{The oracle-integrity auditing workflow.}
The unit under test is the oracle, taken as a program with a four-operation interface
(\textbf{left}): make a fresh environment, take a state signature, call the oracle, and name the
class the oracle lives in. Three families of checks run against that interface (\textbf{centre}):
five state-integrity invariants, three known-answer controls, and a fault-injection battery of seven
defect classes realised by 21 variants across three targets. A differential-probe validity gate
decides which variant--target combinations count as separable \emph{before} any detection number is
computed (\textbf{right}); a combination the probes cannot separate is reported as unresolved and is
never counted as equivalent. The suite detects 54 of the 57 separable combinations; its three misses
are one variant on all three targets, named in Section~\ref{sec:res-oracle} rather than summarised
away.}
\label{fig:workflow}
\end{figure}

The suite requires four operations of any object it audits: make a fresh environment for a given
task, produce a comparable signature of the state being graded, call the oracle, and name the class
the oracle lives in so that a defect can be injected. Nothing in the interface is specific to the
benchmark under study; an adapter still needs access to the environment's state and action record.

On that interface we define eight checks in two families. The \emph{invariants} state properties the
oracle must satisfy whatever the task: \textbf{INV-1} a direct call must not change the graded
state; \textbf{INV-2} two calls on the same environment must agree in verdict and leave the same
state; \textbf{INV-3} the verdict must not depend on the order in which episodes are scored;
\textbf{INV-4} if the oracle raises part-way through, the state must be restored and the exception
must propagate; \textbf{INV-5} after a session-ending step the environment must still hold the
state the agent left, and no actions beyond the terminating one may have been appended. The
\emph{controls} validate the direction of the verdict against a known answer: \textbf{CTL-1} an
episode that accomplishes nothing, and one that says the required outputs without performing the
required writes, must not be graded as solved; \textbf{CTL-2} the reference solution must be graded
as solved; \textbf{CTL-3} an accepted alternative spelling of a required output must be accepted.

To measure the suite's sensitivity and its behaviour under the un-injected condition we inject
seven defect classes into the oracle, one at a time, each constructed around one designated defect
mechanism; some variants violate more than one oracle property (C3, for example, also lacks
exception protection; Section~\ref{sec:res-oracle}): a state mutation that is never restored; a
restore that happens only on the success path; a restore that forgets the recorded actions; a
verdict that depends on an order through a module-level cache; an oracle that always passes; one
that ignores the state component; and one that matches outputs without any normalisation. The
unmodified oracle is the un-injected control.

\subsection{Pre-registered follow-up ablation and boundary probes}\label{sec:p0-method}
A pre-registered follow-up ablation tested the full suite's incremental value, the accuracy of its
upstream fault models, and a boundary of the frozen design. Its protocol (SHA-256
\texttt{8dfb69cf\ldots}) was frozen before the no-model-call run. Four baselines, B1--B4, used the
same probes and task samples as their corresponding checks but simpler criteria: \textbf{B1} compares
only the database hash on the reference-solved probe; \textbf{B2} compares only the action sequence
on that probe; \textbf{B3} scores the same environment twice and compares only the verdicts; and
\textbf{B4} combines the known-answer negative controls with the positive reference control.
\textbf{B5} is their logical OR, hence a structural subset of SimGlot rather than an external method.

The four groups stayed separate: \textbf{E1} contains C1--C7; \textbf{E2} contains the 14 G
variants; \textbf{E3} compares R1 and R2 with C1 and C3 for modelling accuracy; and \textbf{E4}
compares R3-buggy with R3-fixed, which differ only in output matching, to test an out-of-design
boundary. R1--R3 run only on the two $\tau$-bench targets; the synthetic target is N/A. Cells are
\textbf{DETECTED} when a check completes and reports a violation, \textbf{NOT\_DETECTED} when it
completes without an alarm, \textbf{N/A} when inapplicable, and \textbf{ERROR} when the driver raises
an unhandled exception. N/A and ERROR are excluded from both detection and non-detection; a
deliberate INV-4 exception is part of that check's semantics, not a driver ERROR.

Each method $\times$ variant $\times$ target cell ran in a fresh Python subprocess with modules,
variants and samples reinitialised; B5 was derived from B1--B4. The driver first reproduced all 528
SimGlot cells in the frozen matrix. The run produced 936 check-level records across 393 subprocesses
with no failure; timing counts check execution only, with B5 the sum of B1--B4 times.

\subsection{Benchmark, task suite and configuration space}\label{sec:bench}
We use a tool-use benchmark of the ``agent--user--database'' form, in two of its domains: a retail
domain, from which 30 of the test tasks were sampled for the pre-registered matrix, and an airline
domain, whose 50 test tasks are used in full. The airline domain was added in a later round of the
study and required a Chinese version of its task instructions; that translation was produced by a model and
then checked mechanically --- identifiers, ISO dates, flight numbers, airport codes present in the
English text, and every numeral must appear verbatim in the Chinese, and no Latin fragment absent
from the English may appear --- with the residual limitation that semantic equivalence rests on the
mechanical check plus one reading pass, and not on a second independent translator.

Two factors are manipulated. The \emph{user language} is the language in which the simulator
writes. The \emph{environment language} is the language of the three documents the agent reads:
its policy wiki, its tool descriptions, and the strings its tools return on error. Database contents
--- names, addresses, order and product identifiers --- are never translated, because they are data
rather than prompt, and enumerated literals that a tool contract requires verbatim are kept in
English inside the translated text, so that a localised arm cannot fail through argument validation
instead of through language. The environment factor has a per-component form in which exactly one
of the three documents is translated, and a length-padded form in which a semantically redundant
appendix restores the Chinese prompt to the English length. Every arm is admitted by a gate that
asserts, component by component, that the documents claimed to be translated are byte-identical to
the project's single set of translations and that those claimed not to be translated are the
English originals by object identity.

\subsection{Design and pre-registration}\label{sec:design}
The pre-registered matrix varies the user language with the task, the agent, the simulator model and
the repetition held fixed --- 30 tasks $\times$ 2 languages $\times$ 6 simulators $\times$ 4 agents
$\times$ 3 repetitions = 4{,}320 episodes --- and is analysed from the archived episodes rather
than regenerated. A second pre-registered factor crosses the user language with the environment
language, 1{,}080 episodes, plus 1{,}080 in the length-padded environment. Three further
experiments, whose designs were frozen before the corresponding runs, were added in a later round
of the study: a per-component ablation of
the environment factor (3{,}240 episodes), a repeat of the user-language manipulation in a second
domain (1{,}200 episodes), and an independent adjudication of the collapse label (212 archived
episodes).

Two arms were changed after their pilot data were seen, and we report this as a deviation rather
than presenting the final design as the planned one. The second-domain experiment was to include a
cross-vendor agent; its pilot collapsed in none of 120 episodes, so by the pre-registered stopping
rule that arm was dropped, halving the experiment and leaving independent-model validation open.
The adjudication was to draw 200 collapsed episodes; the archive contains 106, so it became a
census. The pilot itself was a directional check with a pre-registered decision rule, not a
significance test.

\subsection{Measurement and statistics}\label{sec:stats}
The unit of randomisation we care about is the task, so the primary inference is a task-clustered
interval: we aggregate the paired difference within each task, take the mean across tasks, and use
the $t(G-1)$ reference distribution, with $G$ the number of tasks. Comparisons are paired within
the same task, simulator, user language and repetition, so that a component arm is compared with
the baseline on cells that differ in exactly the manipulated component. Exact and permutation tests
are reported, where they appear, as cross-checks on an estimate and are never used on their own to
declare an effect resolved. Every reported score is computed under the state-corrected grading
described in Section~\ref{sec:res-oracle}; where an archived arm's labels required reconstruction,
we rebuilt the state component by replay and left the textual component as archived, since the
defect affects only the former. An interval that contains zero is reported as no detected difference.

\subsection{Replication package}\label{sec:replication}
The package accompanying this paper is assembled so that every reported statistic can be recomputed
offline from the released records, on a machine with no network access, no API keys and no model
client; the AppWorld probes and the model interactions are archived, not re-executed. It contains
four parts.

\begin{itemize}
  \item \textbf{The auditing suite and its fault catalogue.} The eight checks
        (Section~\ref{sec:oracle-kit}), the seven injected oracle-defect classes, the positive and
        negative controls, and the frozen map from variant to defect class that the ablations
        aggregate over.
  \item \textbf{The frozen per-episode records.} The v1.0 main matrix, the component ablation and
        the second task domain, together with the archived labels and the state-corrected labels
        produced by the replay described in Section~\ref{sec:res-oracle}. Archived and state-corrected
        values are kept side by side rather than overwritten.
  \item \textbf{The pre-registration and the deviation logs.} The pre-registered design, and a
        numbered log recording every departure from it -- including the two arms dropped on pilot
        evidence, the partition adjustment in the second benchmark, and the two defects found in our
        own validity gate. We record these rather than presenting the final grid as the planned one.
  \item \textbf{Offline entry points.} One command runs the offline analyses in groups and
        prints an exit code per group: the engine-trace adjudication recomputation, the audit
        ablations and baselines, the oracle-integrity regressions, and the source-binding check.
        The entry point reports which source archives it could verify and which were absent; it does
        \emph{not} count an absent archive as verified. The follow-up ablation has its own analysis
        entry point in its ancillary archive (Table~\ref{tab:repro}).
\end{itemize}

Table~\ref{tab:repro} states, for each analysis in this paper, which frozen source backs it, which
command recomputes it, and what that command can and cannot re-establish. The distinction is not
cosmetic. Running the repository's analysis entry points reproduces the analyses, not the
episodes: no entry point re-queries a model, and the adjudicator's raw verdicts cannot be
re-obtained offline --- only their agreement statistics are recomputable from the archived
verdicts. The follow-up ablation ships as a separate ancillary archive whose analysis step recomputes
its matrix and summary offline from the archived raw records.

\begin{table}[t]
\caption{Analysis, frozen source, entry point and scope of reproduction. Commands are run from
the root of the archive named in the second column, except the two v2 scripts, which run from
the root of the v2 analysis package distributed as the ancillary file
\texttt{simglot-v2-analysis.zip} with this preprint. ``Recomputed'' means the reported number is
re-derived from frozen per-episode records by a deterministic script; it does not mean the model
episodes can be regenerated.}
\label{tab:repro}
\footnotesize
\setlength{\tabcolsep}{4pt}
\begin{tabular}{@{}>{\raggedright\arraybackslash}p{0.20\textwidth}>{\raggedright\arraybackslash}p{0.13\textwidth}>{\raggedright\arraybackslash}p{0.28\textwidth}>{\raggedright\arraybackslash}p{0.31\textwidth}@{}}
\toprule
Analysis (section) & Frozen source & Entry point & Recomputed offline \\
\midrule
Archived and state-corrected labels; main matrix (\S\ref{sec:res-oracle}) & v1.0 archive &
\texttt{bash scripts/}\newline\texttt{offline\_repro.sh} & yes: per-episode records, replay and state-corrected labels;
no model client, no key, no network \\
Oracle-integrity matrix; component ablation; cross-domain experiment; adjudication statistics
(\S\ref{sec:res-oracle}--\S\ref{sec:res-general}) & S2 archive &
\texttt{cd sqj \&\& bash}\newline\texttt{swe/reproduce\_s2.sh} & yes for the matrices, intervals and agreement
statistics; the adjudicator's verdicts are archived, not re-queried \\
Fault battery; detector ablations; power analysis; portability attempt
(\S\ref{sec:res-oracle}, \S\ref{sec:threats}) & enhancement archive &
\texttt{bash checks/}\newline\texttt{clean\_room.sh <S2 root>} & yes with the S2 archive present; the AppWorld
probes are not reproducible and its task data is not distributed \\
Label transition and agent comparison (\S\ref{sec:res-oracle}) & v1.0 archive and its
released replay table & \texttt{python3 v2\_01\_}\newline\texttt{label\_transition.py}\newline and \texttt{v2\_02\_agent\_gaps.py}\newline (v2 analysis package) & yes: deterministic from the archived labels and the replay \\
Follow-up ablation and boundary probes (\S\ref{sec:p0-incremental}--\S\ref{sec:p0-boundaries}) &
P0 ancillary archive & \texttt{python3 code/driver.py}\newline\texttt{analyze <ws> out-full} &
yes: matrix, summary, gate and timing from the archived raw records, given the v1.0 frozen
\texttt{b\_matrix.csv} and \texttt{b\_validity.csv} under \texttt{<ws>/repo/results/summary}; the
cell run itself (\texttt{run\_full.sh}) records the executed command with machine-specific paths \\
\bottomrule
\end{tabular}
\end{table}

Two things are deliberately \emph{not} in the package. The second independently implemented
benchmark used for the negative result in Section~\ref{sec:threats} ships its task data under a
licence that forbids plaintext redistribution, so the package carries the adapter, the applicability
findings and the version fingerprint, but not the data. And the model-served episode transcripts
are published in aggregate and in the adjudication inputs, with the simulated user turns replaced by
a placeholder; redistribution is limited to the published records and their stated terms.

A container-free environment file pins the analysis dependencies; the model client is not among
them. Nothing in the package calls a model, and the recomputation of every reported statistic is
deterministic given the frozen records.

\section{Results}\label{sec:results}

\subsection{RQ1 (primary): How reliable is the evaluation oracle?}\label{sec:res-oracle}

\paragraph{The suite and its validation.}
We implemented the eight checks of Section~\ref{sec:oracle-kit} against an interface of four
operations --- make a fresh environment, take a state signature, call the oracle, and name the
class the oracle lives in --- and ran them on three targets: the oracle of a retail-domain agent
benchmark, the oracle of the same benchmark's airline domain, and a synthetic domain written from
scratch. The first two are \emph{two task domains of one scoring implementation}, rather than
independent frameworks; the second domain contributes a different task suite, tool surface and
database against the same oracle code. The synthetic domain
exists to show that the suite is not shaped around one benchmark and to provide a
ground-truth-controlled setting. Throughout, \emph{three targets} means these three.

For each target we injected seven defect classes, one at a time, into the oracle. Table~\ref{tab:kit}
gives the outcome.

Two summaries of the same run are possible, and they differ. At the level of the seven defect
classes, every class is detected on every target. The battery is finer than that: each class is
realised by several concrete variants of the reward function, 21 variants in all, and running them
against three targets gives 63 variant--target combinations. On the \emph{primary denominator}
--- the 57 combinations that a differential probe showed to be behaviourally distinct from the
shipped implementation, the remaining 6 being left unresolved by the probes and never counted as
equivalent --- the full suite detects \textbf{54 of 57}. On the \emph{conservative denominator},
which also counts the 6 unresolved combinations, it detects 56 of 63. We report both and do not
mix them: they answer different questions, and the larger number is not the stronger result. The
headline of this subsection is the smaller number, because it is the one a reader can check
against the released fault matrix.

The identity of the un-injected baseline matters, and we state it explicitly because the two
sentences that follow would otherwise appear to contradict each other. The implementation this suite
audits is the copy the package ships, and that copy carries the repair described below; the class
labelled \emph{as shipped} in Table~\ref{tab:kit} is therefore the \emph{repaired} implementation,
and it passes every check on every target. The class labelled \emph{state mutation} is not a
hypothetical: it is a faithful reproduction of the same function \emph{before} the repair, with no
restore statement of any kind. Against that reproduction, INV-5 fires on every task we sampled in
both domains. A three-way check confirms the separation --- the vendored source contains the
snapshot-and-restore, the injected variant contains no restore at all, and INV-5 is silent on the
first and alarming on the second in both domains --- and it is released with the package. The suite would therefore have caught the defect had it been pointed at the benchmark as it stood
before the repair; the claim is that it catches one of the seven classes, not that it fires on the
code we ship.

\paragraph{The miss, and why a class-level summary hides it.}
Table~\ref{tab:kit} summarises the injection experiment at the level of the seven defect classes,
which is how the experiment was designed --- and it is more generous than the result. The suite is
therefore \emph{not} free of blind spots, and the miss is the same variant,
\texttt{G5b\_exception\_swallow}, on all three targets: an oracle that swallows an exception raised
inside itself, so that INV-4 records the
probed path as \emph{not applicable} rather than as a violation. The mechanism is worth naming
because it is not an accident of this implementation --- a check whose precondition is destroyed by
the defect it is meant to detect will record that defect as inapplicable. We report the miss rather
than the class-level summary that conceals it.

\paragraph{Which checks are load-bearing.}
The last column of Table~\ref{tab:kit} is computed rather than asserted. A check is \emph{necessary}
for a class when removing it leaves some detected combination of that class undetected by any
remaining check; \emph{none} means no single check is necessary, because two or more of them
independently cover the class. Four of the eight checks are necessary for at least one class and
four are not for any: INV-3 (9 unique detections across the battery), CTL-1 (15), INV-4 (6) and
CTL-3 (5), against INV-1, INV-2, INV-5 and CTL-2 with zero. Redundancy is not a flaw in a test suite
--- an independent check that agrees is how a suite earns trust --- but it does mean the eight
checks are not eight independent capabilities, and the detector ablation in the replication package
is computed on that basis. The two families remain non-interchangeable: an oracle that is
state-clean but wrong is invisible to the invariants, and one that is state-dirty but accidentally
right is invisible to the controls.

\begin{table}[t]
\caption{Defect class $\times$ detector, computed from the released fault matrix over all three
targets. The second column counts \emph{separable variant--target combinations}, not variants: the
battery has 21 variants and three targets, giving 63 combinations, of which the differential probe
resolved 57, distributed across the classes as shown. \textbf{Fires} lists every check that alarmed
on at least one variant of that class;
\textbf{necessary} is the check whose removal leaves some detected variant of that class undetected
by any remaining check. Variants within a class do not all produce the same alarm pattern, which is
why the table reports a union rather than a single pattern; the class \emph{saturated oracle}
contains one variant, on each target, that no check alarms on at all
(Section~\ref{sec:res-oracle}). Every variant is author-constructed, so these figures are a
conditional property of this battery, not a detection rate for unknown defects.}
\label{tab:kit}
\begin{tabular}{@{}llll@{}}
\toprule
Defect class & separable comb. & checks that fire & necessary \\
\midrule
state mutation & 9 & \texttt{INV\text{-}1}, \texttt{INV\text{-}2}, \texttt{INV\text{-}4}, \texttt{INV\text{-}5}, \texttt{CTL\text{-}1}, \texttt{CTL\text{-}2} & \texttt{CTL\text{-}1} \\
exception leak & 9 & \texttt{INV\text{-}1}, \texttt{INV\text{-}2}, \texttt{INV\text{-}4}, \texttt{INV\text{-}5}, \texttt{CTL\text{-}2}, \texttt{CTL\text{-}3} & \texttt{INV\text{-}4} \\
actions not restored & 9 & \texttt{INV\text{-}1}, \texttt{INV\text{-}2}, \texttt{INV\text{-}4}, \texttt{INV\text{-}5} & \emph{none} \\
order dependence & 9 & \texttt{INV\text{-}3} & \texttt{INV\text{-}3} \\
saturated oracle & 9 & \texttt{CTL\text{-}1} & \texttt{CTL\text{-}1} \\
database-blind oracle & 7 & \texttt{CTL\text{-}1} & \texttt{CTL\text{-}1} \\
strict output match & 5 & \texttt{CTL\text{-}3} & \texttt{CTL\text{-}3} \\
\bottomrule
\end{tabular}
\end{table}

\paragraph{What the suite found on the benchmark.}
The benchmark's reward function resets the environment database in order to replay the reference
solution and construct a ground-truth hash, and it never restores the database afterwards. The
first returned reward is unaffected by this particular state-contamination mechanism, because the
agent's database hash is captured before the reset, and the benchmark's own agent consumes that
value without reading the environment again. What turns the side effect into a wrong label is a
harness that scores the episode separately, after the session has ended: the second read finds the
reference solution's state rather than the agent's. Our harness did exactly that.

Invariant INV-5, terminal-state stability, fires on that reproduction for this reason, and it
fires without any downstream code being involved: after a session-ending step the environment no
longer holds the state the agent left. The archived episodes analysed below use the unpatched
implementation, and the injection experiment above deliberately reproduces the same unpatched
logic to measure detector sensitivity. INV-5 raises no alarm on the repaired copy we ship (no alarm
under the un-injected condition). The counter-example is minimal. We construct an episode in which
the agent performs no write operation at all and simply says goodbye; on the archived code path
this episode is graded as a database pass, and after the repair it is not. C3 restores data only on
the success path; when reward calculation raises, the environment is left un-restored, so INV-4
alarms in addition to INV-1, INV-2 and INV-5, as Table~\ref{tab:kit} records.

\paragraph{The correction.}
Because the benchmark exposes no deterministic seed and the provider does not offer immutable model
versions, the 4{,}320 archived episodes cannot be regenerated; they can, however, be \emph{replayed}.
We re-execute each archived tool call against a clean database and compare the resulting terminal
hash with the reference hash. The replay input is the \texttt{agent\_trace} recorded step by step in
the agent's own message loop, not \texttt{env.actions}, so the gold-action suffix documented in
upstream issue~\#92 cannot enter this replay \citep{taubench_issue92}. All 137 episodes that ended by
hitting the step limit --- which never pass through the defective path, and whose database labels
were therefore unaffected by it --- reproduce their archived labels exactly. All 33{,}402 tool
observations reproduce the archived strings character for character, with no mismatch. Observation
agreement supports, but does not by itself prove, identity of hidden state.

Under the state-corrected labels the archived 4{,}194 database passes become 2{,}883, and overall
success falls from 94.3\% to 65.5\%. The correction removes the identified database-state
contamination but retains the original output-matching semantics, which
Section~\ref{sec:p0-boundaries} shows to be over-broad; it is not a fully validated semantic success
rate. Every number reported below is computed under the state correction.

\paragraph{What the correction does to the comparison between systems.}
An inflated score and a corrupted comparison are different failures, and the archive separates
them. The correction is not a constant offset. Of the 4{,}320 episodes, 2{,}829 pass under both
labels and 1{,}244 pass only under the archived one; no episode moves in the other direction, which
is what a defect that reads the reference solution's state should produce. The size of the change
differs by agent, from 24.0 to 33.6 pp, and every one of those intervals excludes zero
(Table~\ref{tab:agentlabels}).

\begin{table}[t]
\caption{Archived and state-corrected success by agent (4{,}320 episodes, 1{,}080 per agent).
$\Delta$ is the change from the archived to the state-corrected label, with a task-clustered 95\%
interval (G${=}$30, $t(29)$); ranks are within this matrix. The state-corrected comparison is
conditional on the 30 sampled tasks and on a single serving provider, and inherits every limitation
in Section~\ref{sec:threats}.}
\label{tab:agentlabels}
\footnotesize
\setlength{\tabcolsep}{3pt}
\begin{tabular}{@{}lrrrrr@{}}
\toprule
Agent & Archived & Corrected & $\Delta$ pp [95\% CI] & Rank (arch.) & Rank (corr.) \\
\midrule
DeepSeek-V4.1-Flash & 98.0\% & 64.4\% & $-33.61$ $[-46.86,-20.36]$ & 1 & 3 \\
Qwen3.7-Plus        & 97.7\% & \textbf{73.1\%} & $-24.54$ $[-36.27,-12.80]$ & 2 & 1 \\
Qwen3.8-Max         & 95.9\% & 71.9\% & $-23.98$ $[-34.39,-13.57]$ & 3 & 2 \\
Qwen3.8-Flash       & 85.6\% & 52.5\% & $-33.06$ $[-41.51,-24.60]$ & 4 & 4 \\
\midrule
Pooled              & 94.3\% & 65.5\% & $-28.80$ $[-38.79,-18.81]$ & --- & --- \\
\bottomrule
\end{tabular}
\end{table}

Under the archived labels the highest point estimate is DeepSeek-V4.1-Flash at 98.0\%, but its margin
over Qwen3.7-Plus is 0.28 pp with an interval of $[-0.82,+1.38]$ that contains zero: the archived
comparison never resolved a leader. Under the state-corrected labels Qwen3.7-Plus has the highest
point estimate at 73.1\%,
8.80 pp above DeepSeek-V4.1-Flash $[1.98,15.62]$; Qwen3.8-Max is second, 7.59 pp above it
$[0.53,14.65]$; and the correction removes 9.07 pp more from DeepSeek-V4.1-Flash than from
Qwen3.7-Plus ($[2.49,15.66]$; difference-in-differences on cells matched by task, simulator, user
language and repetition). This is a statement about the labels, not a ranking of the models: the
four agents are served through one provider, the tasks are the 30 sampled tasks, and every
limitation in Section~\ref{sec:threats} applies. The state-corrected top two are not separated
either:
Qwen3.7-Plus leads Qwen3.8-Max by 1.20 pp $[-3.54,+5.95]$, an interval that contains zero. The
correction changes which systems form the leading group --- Qwen3.7-Plus and Qwen3.8-Max above
DeepSeek-V4.1-Flash --- and leaves a single winner unresolved. What it shows is that the defect changes the
answer to the question a benchmark exists to answer --- which system is better --- and an inflated
absolute score alone would not do that.

The spread across tasks carries the same point. 29 of the 30 tasks lose at least one archived pass;
the median task loses 19.1 pp and nine tasks lose more than 40 pp. A correction reported only as a
pooled drop from 94.3\% to 65.5\% therefore understates it: the score moved by 28.8 pp, but not
uniformly, and the comparisons the benchmark supports move with it. Both the transition matrix and
the per-agent comparison are recomputed by the v2 analysis package from the archived and
state-corrected labels (Table~\ref{tab:repro}).

\subsubsection{Incremental detection value of the full suite}\label{sec:p0-incremental}
The pre-registered follow-up ablation compares the full suite with B5, a basic-check suite
constructed from its own components. Table~\ref{tab:p0} reports each group separately and uses the
target--variant combination as the statistical unit. Beyond basic state checks and known-answer
controls, the additional checks detected 20 target--variant combinations on the primary
denominator: eight in E1 and 12 in E2. Of these, INV-3 uniquely detected nine (C4, G4a and G4b
on each of three targets), INV-4 uniquely detected six (C2 and G2b on each target), and CTL-3
uniquely detected five (C7 on two targets, G7a on two, and G7b on one). Each of the 20 was signalled
by exactly one check, and that check matches the property injected; INV-5 made no unique
contribution. The three E2 misses are the already reported \texttt{G5b\_exception\_swallow} on the
three targets.

\begin{table}[t]
\caption{Follow-up ablation detection counts by pre-registered group. The unit is a target
$\times$ variant combination. \emph{Primary} excludes the six unresolved combinations; \emph{all}
includes them. B1--B4 are reported on the primary denominator, as pre-registered; B5 is their logical
OR and is a structural subset of SimGlot. E1 and E2 are not pooled into a new detection rate.}
\label{tab:p0}
\footnotesize
\setlength{\tabcolsep}{2.5pt}
\begin{tabular}{@{}llrrrrrr@{}}
\toprule
Group & Denominator & SimGlot & B1 & B2 & B3 & B4 & B5 \\
\midrule
E1 (C1--C7) & primary 20 & \textbf{20/20} & 0 & 6 & 0 & 9 & 12/20 \\
E1 (C1--C7) & all 21 & \textbf{21/21} & --- & --- & --- & --- & 12/21 \\
E2 (14 G) & primary 37 & \textbf{34/37} & 4 & 12 & 0 & 14 & 22/37 \\
E2 (14 G) & all 42 & \textbf{35/42} & --- & --- & --- & --- & 22/42 \\
\bottomrule
\end{tabular}
\end{table}

This is a suite ablation, not a comparison with an external method: B5 is contained in SimGlot by
construction, so the containment itself is not an empirical finding. The informative result is
where the additional detections fall. B1's low count reflects its reference-solved probe, on which
the reference replay leaves the same database state the agent already produced; it does not show
that database-hash comparison is ineffective under other probes. B3 detects none of the 57 primary
combinations because it compares verdicts without comparing post-scoring state. On check execution
time alone, the full suite costs 1.72$\times$ B5 on retail (654.2 vs 380.7 seconds) and 1.78$\times$ on
airline (1{,}493.7 vs 837.5 seconds).

\subsubsection{Validation against upstream implementations}\label{sec:p0-upstream}
E3 compares two upstream implementations with their constructed fault models on the two
$\tau$-bench targets. R1 (upstream \texttt{59a200c}) and C1 have identical detection signatures under
the evaluated probes; R2 (PR~\#89 with data preservation enabled) and C3 also have identical
signatures. Across the 12 checks, both comparisons agree on every check in both targets. These
results support modelling accuracy and are not counted as independent defect detections.

The seven pre-registered predictions covered no alarm on C0, the classes B5 would miss, the E2
detection gap, the R1/C1 and R2/C3 correspondences, the R3 result, and zero ERRORs. Six of the seven
pre-registered predictions held; the exception was P5. R2 was expected to add an INV-4 alarm
relative to C3, but no difference appeared. C3 itself lacks exception protection, so its exception
path also leaves data un-restored and INV-4 alarms. The prediction failed because the fault taxonomy
is not fully orthogonal, not because the upstream exception leak went undetected.

\subsubsection{Out-of-design defects and detection boundaries}\label{sec:p0-boundaries}
E4 isolates output matching. R3-buggy and R3-fixed differ only in output matching: R3-fixed applies
the word-boundary match of PR~\#89 to every required output. The fixture and its specification cover
only the numeric case --- a required number appearing inside a longer digit string. Every method has
identical marks on both conditions, and R3-buggy has the same marks as C0. All 27 fixture probes
behave as specified: prefix-digit and suffix-zero near-misses are accepted by R3-buggy and rejected by
R3-fixed, while exact answers pass both. The frozen suite has no negative control for an approximate
but incorrect output, so no check detects this boundary under the evaluated probes. Together with
\texttt{G5b\_exception\_swallow}, R3 defines the reported limits; no detector was added after the
result.

\subsection{RQ2 (secondary): How do harness configurations influence observed evaluation outcomes?}\label{sec:res-config}

The agent's own world is written in three documents: the policy it reads, the descriptions of its
tools, and the strings its tools return on error. Manipulating all three at once tells us that the
world matters but not which part of it; we therefore ran each component alone. Three arms
localise the policy, the tool descriptions, or the error strings into Chinese and hold the other
two in English, over 30 tasks $\times$ 6 simulators $\times$ 2 user languages $\times$ 3
repetitions = 3{,}240 episodes, with no failed cell. Two archived arms --- the fully English
baseline and the fully Chinese environment --- are reused rather than re-run, and every comparison
is paired within the same (task, simulator, user language, repetition) cell and clustered by task.

Table~\ref{tab:e1} gives the result. All five Chinese arms have intervals that exclude zero against
the English baseline, and the point estimates order the three components with the tool descriptions
largest, the policy next and the error strings smallest. That ordering is descriptive: of the three component-versus-all-three
contrasts we test, two exclude zero and the third does not (policy alone against all three,
$+0.83$\,pp $[-0.36,+2.03]$). No other pairwise comparison is reported as resolved, and we read
none of them as identifying a mechanism.

\begin{table}[t]
\caption{Component ablation of the environment localisation (retail; 1{,}080 paired cells per arm,
except the length-padded arm, whose corrected table carries 1{,}079 --- one cell failed at run time
and has no transcript, see the Length paragraph).
$\Delta$ is the paired difference against the fully English baseline, with a task-clustered 95\%
interval (G${=}$30, $t(29)$). Success differences are computed under the state-corrected scorer.}
\label{tab:e1}
\setlength{\tabcolsep}{2pt}
\begin{tabular}{@{}lrrr@{}}
\toprule
Arm & Collapse & $\Delta$ collapse & $\Delta$ success \\
\midrule
English baseline            & 9.81\% (106/1080) & --- & --- \\
Tool descriptions Chinese    & \textbf{0.65\%} (7/1080) & ${-}9.17$ $[-11.61,-6.73]$ & $+9.63$ $[+5.09,+14.17]$ \\
Chinese environment (all three) & 1.76\% (19/1080) & $-8.06$ $[-10.57,-5.54]$ & $+5.56$ $[+0.45,+10.67]$ \\
Chinese $+$ length padding   & 1.30\% (14/1079) & $-8.52$ $[-10.77,-6.28]$ & $+5.80$ $[+0.56,+11.04]$ \\
Policy Chinese               & 2.59\% (28/1080) & ${-}7.22$ $[-9.46,-4.98]$ & $+5.09$ $[+0.80,+9.38]$ \\
Error strings Chinese        & 4.54\% (49/1080) & ${-}5.28$ $[-8.03,-2.52]$ & $+5.19$ $[+1.31,+9.06]$ \\
\bottomrule
\end{tabular}
\end{table}

Two claims follow, and both are narrow:

First, \textbf{all three components are associated with a lower collapse rate when localised
alone, and the effects do not combine additively}. The
weakest single arm still moves collapse by $-5.28$\,pp. Each component is therefore associated
with a lower measured collapse rate in the tested configuration when it is localised on its own;
but these single-component contrasts do not establish that any of the three is \emph{necessary}.
The three component arms share one batch, but their baseline comes from a separate batch, so even
this association is not a within-session causal
estimate (Section~\ref{sec:threats}). The strongest single arm (tool descriptions, $-9.17$\,pp) also has a larger point
estimate than the arm that changes all three documents ($-8.06$\,pp): the full manipulation does
not beat its own largest component, and localising only the tool descriptions is measured
$1.11$\,pp above it $[0.36,1.86]$. Symmetrically, localising only the error strings is measured
$2.78$\,pp below the full manipulation $[1.21,4.34]$, and the policy alone does not differ from it
($+0.83$\,pp $[-0.36,+2.03]$).

The observed pattern is descriptive rather than a ranking of the three documents. The contrasts
come from different run periods (Section~\ref{sec:threats}); the
pre-registration set no hypothesis on differences between components, and the intervals carry no
multiplicity correction; and our own power analysis puts a contrast of $1.11$\,pp at \textbf{17\%}
power in this design, against 85\% at $4$\,pp and 100\% at $9$\,pp. An interval that excludes zero
is evidence that the contrast in these data is not zero. It is not evidence that the design can
resolve a difference of this size against batch drift, and a low-power design that reaches
significance tends to overstate the effect it has found. What the design supports is therefore
narrower than a ranking: each component changes the measured outcome when localised alone, the
effects do not combine additively, and the relative size of the three is an observed ordering in
this configuration whose mechanism is not established. A same-period replication with the arm order
reversed would reduce this batch confound, and we did not run one.

We describe the pattern as a \emph{non-additive observed pattern} rather than as sub-additivity,
because with only the single-component arms and the all-three arm we cannot identify the
interaction terms of higher order; what the design supports is the observation that the components
do not combine as independent contributions.

Second, the error strings are the \emph{weakest} component. We had pre-registered the opposite:
the mechanism hypothesis located the failure at the tool-error boundary --- the identity check that
returns \texttt{user not found} --- and therefore predicted that the error vocabulary would carry the
effect. It carries the least of the three. We report this disagreement between the pre-registered
hypothesis and the observed result directly.

\paragraph{Length.}
Localising a document also changes its length, and for Chinese it shortens it: measured in
characters, the Chinese policy wiki is 36\% of the English length, the tool descriptions 66\%, and
the error strings 30\%. A length-matched fourth environment, padded back to the English prompt
length, moves collapse by $-0.46$\,pp and success by $+0.24$\,pp relative to the unpadded Chinese
environment --- one order of magnitude below any component effect. We detect no length effect at
this resolution, and length alone does not account for the component effects; the interval for the
padded arm is reported with the other arms and crosses zero. Length
is not, however, fully excluded for the single-component arms: the
padding controls the length of the environment as a whole, and we did not build a length-matched
counterpart of each individual document. A per-component length confound of the observed size would
have to be an order of magnitude larger than the whole-environment length effect to overturn the
ranking, which we regard as implausible but do not rule out. The interval for the padded-versus-unpadded
Chinese contrast itself is $-0.46$\,pp $[-1.39,+0.56]$ and contains zero; what Table~\ref{tab:e1}
reports for the padded arm is that arm against the English baseline, and those intervals exclude zero
because the language effect is present in both Chinese arms.

One bookkeeping difference between the arms belongs here. The length-padded arm carries 1{,}079 of
its 1{,}080 cells: task~27 with the qwen3.8-flash agent, the deepseek-v4.1-flash simulator, Chinese
user language and repetition~0 failed at run time and has no transcript, so it is absent from the
state-corrected per-episode table from which this arm's figures are computed. Treating that cell as
a failure moves the arm's collapse contrast from $-8.52$ to $-8.53$\,pp and leaves every conclusion
in this section unchanged; the two-sided sensitivity is reported with the length-matching analysis
in the package.

\paragraph{A second domain reverses the sign.}
We repeated the user-language manipulation on the benchmark's airline domain --- a different task
suite, a different tool surface, and a substantially harder set of tasks --- over 50 tasks
$\times$ 6 simulators $\times$ 2 user languages $\times$ 2 repetitions = 1{,}200 episodes, with no
failed cell. The domain is where we had hoped to give the stronger agents a non-zero base rate: it
does not do that (the cross-vendor agent we had planned to use collapsed in none of 120 pilot
episodes, so that arm was dropped and independent-model validation remains open, see
Section~\ref{sec:threats}).

The agent whose base rate is non-zero does collapse in the second domain, at 9.67\% (English
simulator) and 4.33\% (Chinese simulator). The direction, however, is the opposite of the one we
measured before:

\begin{center}
\begin{tabular}{@{}llrr@{}}
\toprule
Domain & & English simulator & Chinese simulator \\
\midrule
retail (archived)   & collapse & 6.48\% & 13.15\% \\
airline (this study) & collapse & 9.67\% & 4.33\% \\
\bottomrule
\end{tabular}
\end{center}

The retail figures are per user-language arm of 540 episodes for that agent --- 1{,}080 is the two
arms combined --- and the airline arm is 600 paired cells per language. The airline difference is
$-5.33$\,pp with a task-clustered interval of $[-8.15, -2.51]$, which excludes zero; the retail
difference is $+6.67$\,pp. Reporting two intervals separately does not establish that the two
domains differ, so we also test the contrast directly: the difference between the two effects is
$-12.00$\,pp with an interval of $[-16.33, -7.67]$ (Welch--Satterthwaite, $df = 65.76$; exact $t$ quantile), which excludes zero.
The signs are opposite, each is resolved, and so is the difference between them. We pre-registered
the two ways this could be read, and this is the second: the effect does not generalise, so we
report it as a boundary and restrict the language effect to the retail configuration in which it
was measured. On the success
side the airline difference is $-2.00$\,pp $[-7.05, +3.05]$, which excludes nothing and supports no
direction.

This is a negative result and we report it as one, and we keep it to what two domains can support.
The sign is not stable across the domains we ran, so no direction can be attributed to the
simulator's language; the effect is a property of the configuration in which it is measured. That
is weaker than a mechanism claim: two domains, each changing its task set, tool surface and base
rate together, cannot separate a configuration property from an interaction between language and
domain. The reversal also rules out a direction attributed to Chinese itself; the measured sign
belongs to the evaluated configuration.

\subsection{RQ3 (secondary): How sensitive are the conclusions to execution factors, and where do they stop holding?}\label{sec:res-general}

RQ1 and RQ2 report what the audit found. RQ3 reports where each of those answers stops holding:
the \emph{implementation} boundary, at which the same auditing method failed to transfer to a
second, independently implemented framework; and the \emph{instrument} boundary, at which the
criterion this paper relies on turns out to be an operational definition corroborated by a language
model rather than a human ground truth. We report both as results rather than as omissions. The
task domain is itself part of the configuration, and we treat its effect under RQ2 for that reason.

\paragraph{The implementation boundary.}
We applied the same oracle-integrity suite to a second, independently implemented agent framework.
The attempt did not yield a validation. The reference terminal state that the positive control needs
cannot be reconstructed by an external harness from that framework's recorded API-call log, so no
sensitivity or specificity could be estimated on its held-out task set and we report none. The
fixture the framework does expose cannot serve as a positive control either: of its 21 held-out
tasks, the three that look usable are false positives produced by the same state-reconstruction
defect the control exists to detect. The constraint is contractual rather than incidental --- the
framework materialises a reference episode only inside its own run loop --- and we state it as the
precondition a framework must meet before this class of auditing can be ported to it.
Section~\ref{sec:threats} records it as the external-validity boundary it is.

\paragraph{An independent language-model adjudication reproduces the collapse label.}
The collapse criterion is a mechanical operational definition, and a reader is entitled to ask
whether an independent judge would call the same runs degenerate. We put every collapsed episode in
the archive (106) together with a matched non-collapsed control, masked the language arm and the
criterion's own intermediate quantities by replacing each user turn with a placeholder, and had a
judge from a different model family decide, blind to the hypothesis, whether the agent stops making
progress. Of 212 items the judge returned a parseable verdict on 210 (0.9\% attrition). Agreement
with the mechanical criterion is 98.1\% (Wilson 95\% $[95.2, 99.3]$), Cohen's $\kappa$ is 0.962,
sensitivity is 98.1\% and specificity 98.1\%.

The quantity that matters for this paper is whether the agreement differs between the language
arms: 98.6\% in the English arm (one discordant item) and 97.9\% in the Chinese arm (three
discordant items). We did not observe a clear difference. That is weaker than independence, and
the distinction is what this subsection is about. Agreement is at the ceiling, so the comparison
has little power to detect a moderate arm-dependent bias; the 212 items are the archive's positives
plus matched negatives rather than a draw from a deployment distribution, so the reported
sensitivity and specificity are properties of that selected sample and do not transfer to a
population with a different base rate; and the residual leakage is real --- to judge whether the
agent stops making progress the judge must read what the agent said, and in the Chinese arm those
words are Chinese. We therefore claim that no clear dependence of the criterion's agreement on the
arm was observed, not that independence was established, and not that the judge is blind to
language. The judge is also a language model, not a person; this study has no annotator budget, so
the criterion is reported as an operational definition with independent corroboration rather than
as a natural kind.

\section{Discussion}\label{sec:discussion}

\paragraph{An evaluation oracle is software, and it needs the tests other software gets.}
The defect we report is not exotic. A function that mutates the state it is grading and returns
without restoring it is the ordinary shape of a resource-handling bug; what makes it consequential
here is that the function is the grader, and that its caller had no reason to suspect it. The
mechanism is a \emph{missing} restore, not a partial one: the function resets the database in order
to replay the reference solution, builds the ground-truth hash from it, and returns with the
reference solution's writes still in place. A restore that happens on the success path but not on
the failure path is a different defect; it is one of the injected variants in our battery (the
class we label \emph{exception leak}), and it is not the defect the benchmark carried. The
repeat-scoring check fails on an episode the agent did not solve when the database is the only failing
component: the first call moves the environment to the reference state, so the second call's database
check passes. On a reference-solved episode the two verdicts and the database hashes can agree while
the recorded action sequence changes; INV-2 compares a state signature that includes the recorded
actions, and that is what reveals the defect (Section~\ref{sec:res-oracle}). The terminal-state check
--- after a session-ending step, assert that the environment still holds the state the agent left ---
fails without any downstream code being involved. Neither requires a reference implementation, a
model call, or a human. We suggest that both belong in the continuous integration of any harness that
grades an agent episode, next to the regression test that pins the fix.

\paragraph{What a cross-lingual score table does not say.}
A score table reports a measurement. It does not report the trajectory-collapse rate, and our
results are a case in which the trajectory-collapse rate and the measurement move independently:
in the retail configuration the aggregate score difference is unresolved while the rate at which
one agent's trajectory degenerates roughly doubles, and in the airline configuration the same
manipulation moves the degeneration rate in the opposite direction. A practitioner comparing two
language configurations of the same evaluation therefore cannot read the comparison as a statement
about the model, and cannot even assume the direction of the configuration effect is stable across
task suites. This is a claim about attribution, not about the oracle: a change of language can
alter what the agent does for reasons that have nothing to do with a defective grader, and we do
not infer a grader defect from a score that moved (Section~\ref{sec:rw-flaky}). What makes the
comparison uninterpretable is that the score is attributed to the model while the apparatus that
produced it changed. Reporting a trajectory-collapse rate alongside the score, and checking it in
more than one task suite before acting on it, are the two actions that follow.

\paragraph{Localisation is a configuration change, and it should be tested as one.}
The practical reading of the component ablation is not that one document matters and the others do
not; it is that changing the localisation of an agent's world changes the reliability of the
measurement, and that the change is spread over the parts of the world the agent reads. A team that
localises the user-facing side of an agent while leaving its policy, its tool descriptions and its
error vocabulary in English is running a configuration whose measurement properties differ from
both the fully English and the fully localised ones. Testing that configuration --- rather than
assuming it is an intermediate point between the two --- is what the ablation suggests.

\paragraph{Replay as the audit primitive.}
The correction we report is not a re-run: the provider offers no deterministic seed and no
immutable model version, so the episodes cannot be regenerated. What can be regenerated is the
\emph{effect} of the agent's recorded actions on the state the oracle grades. Replaying the agent's
own recorded tool calls against a clean database reproduces every recorded observation and every
label the defect could not reach; it is what turned an inflated label into a measurable one. The
precondition is narrow and we state it as a precondition rather than as a prediction: the replay
needs a benchmark that records the agent's tool calls in a form an external harness can re-execute,
against a database the harness can reset.
We evaluated it on one such benchmark, and Section~\ref{sec:threats} reports the independent
implementation on which the same precondition did not hold. Within that precondition it is cheap:
the replay is deterministic, needs no model, and can be cached.

\paragraph{Reporting a negative result, and reporting a disagreement.}
Two of this paper's findings are negative: a pre-registered mechanism hypothesis was refuted by the
ablation, and the language effect changed sign in a second task domain. We report both directly:
the first as a disagreement between the pre-registered hypothesis and the observed data, and the
second as a boundary on the configuration in which the effect holds. For a paper whose subject is
the trustworthiness of measurements, this is the consistent position.

\section{Threats to Validity}\label{sec:threats}

\subsection{Construct validity}
\paragraph{What RQ2 and RQ3 are not.}
We state the boundary of the two secondary questions explicitly, because it is the reading a reader
is most likely to reach for. RQ2 and RQ3 are \emph{not} measurements of intrinsic language-model
capability or behaviour. The models are the materials of the study, not its object. What those
questions characterise is the measurement process on an audited pipeline: how design choices in the
harness expose, amplify or obscure variability that is present while a benchmark runs, and where the
resulting conclusions stop holding. The consequence for the evidence is stated in the Introduction
and carried through the paper: RQ1 is direct evidence about the evaluation system, and RQ2 and RQ3
are reported as observations and boundaries at correspondingly weaker strength. The conclusions
are about the measurement process, not about intrinsic model behaviour under localisation.

Collapse is an operational definition: a maximal run of at least five consecutive reply steps each
spending at most two completion tokens, with no tool call. It is mechanical and reproducible, and it
is not a natural kind. The independent adjudication in Section~\ref{sec:res-general} corroborates
that the definition tracks what a reader would call a stalled trajectory, and finds no clear
difference in agreement between the language arms; the judge is a language model rather than a
person, agreement is at the ceiling, and the sample is selected, so neither the corroboration nor
the absence of an arm difference is conclusive. The criterion is also defined on the agent's token counts, which are
not a language-neutral unit of text; the robustness analysis in the replication package shows
that the gap survives character-count, repetition and no-length-threshold alternatives. The token
threshold is therefore treated as a criterion of the evaluated configuration, not as a directly
comparable unit across languages. What we measure is
therefore a property of the evaluated agent--simulator \emph{configuration} --- how often, under
this configuration, a trajectory degenerates --- and not a property of the model alone.

\paragraph{Follow-up-specific validity limits.}
\begin{itemize}
  \item B5 is a logical OR of weakened versions of SimGlot's own checks. The full suite contains it
        by construction, so the 20 additional detections locate the contribution of INV-3, INV-4 and
        CTL-3; they are not evidence of an advantage over an external method.
  \item B1 uses the reference-solved probe from INV-1. On that probe the reference replay leaves the
        database in the state the agent already produced, so B1's low count is probe-dependent and
        does not establish that database-hash comparison is generally ineffective.
  \item The fault taxonomy is not fully orthogonal. C3 violates action restoration and exception-path
        data restoration, which is why the pre-registered P5 prediction failed; R2 and C3 have the
        same evaluated signatures.
  \item Six combinations remain unresolved rather than equivalent. Two of them, retail C7 and G7a,
        are signalled by CTL-3, showing that the differential probes can be less sensitive than the
        detectors; this is also why the all-combination E1 count is 21/21 although its primary count
        is 20/20.
  \item The P7/P8 order probes were corrected once after an external review found that the first
        version cleared the shared cache between scorings and masked order dependence. The corrected
        probes were compared with the detection matrix, then frozen before the follow-up run; they do
        not import the detectors, but their development was not wholly independent of it.
\end{itemize}

\subsection{Internal validity}
The provider exposes no effective deterministic seed: at temperature zero the same request gives
different trajectories, so the three ``repetitions'' per cell are a repetition index and not a
controlled randomisation, and a single run is not evidence. All comparisons are therefore paired
within a cell and reported with task-clustered intervals. A further internal-validity cost is that
the component ablation and the second domain ran on a platform that rebuilds the execution
environment periodically; the ablation batch consequently spans one such boundary. The elapsed gap
was under an hour, and a drift probe run over intervals an order of magnitude longer than that gap
found no evidence of version drift, but we cannot exclude it.

A larger instance of the same threat attaches to the component contrasts. The three
single-component arms were run in a single batch, but the arm that localises all three documents
and the fully English baseline come from earlier batches of this project; the gap between the run
periods is at least a day, and the provider exposes no immutable model version, so a provider-side
change between the periods is not excluded. Pairing on task, simulator, user language and repetition
removes task and configuration variance, but it is not a within-session randomisation and cannot
remove drift common to a period. We state the consequence precisely. The \emph{existence} of an
association between each component and a lower collapse rate is supported; the \emph{ordering} of the components should be
read as an observation in this configuration rather than as an established mechanism, and the
$1.11$\,pp contrast between localising the tool descriptions alone and localising all three is the
most exposed of the contrasts we report. A same-period replication with the arm order reversed
would reduce this batch confound, and we did not run one.

Two arms of the second experiment were
dropped on the evidence of their pilot (Section~\ref{sec:res-config}), which is a design decision
made after seeing data, and we record it as such rather than presenting the final grid as the
planned one.

\subsection{External validity}
The auditing method has a portability boundary that we measured rather than assumed. We applied the
same suite to a second, independently implemented agent benchmark. The attempt is a \emph{negative}
result: the reference terminal state needed by the positive control could not be reconstructed from
the framework's recorded API-call log, so no detection rate could be estimated on its held-out task
set, and we report none. The constraint is contractual rather than incidental --- the framework
materialises a reference episode only inside its own run loop --- and we state it as the
precondition a framework must meet before this class of auditing can be ported to it. This bounds
Contribution~1 to the implementations we actually evaluated.

The second benchmark is also one on which a record-level evidence audit has been reported
\citep{gazhou2026bounds}, and the two results do not conflict. That audit asks whether the evidence
stored for a completed run supports the score the benchmark reports, and it can proceed from
records the framework already writes. Ours needs a reference terminal state that an external
harness can construct and compare against, because a positive control must know what the correct
final state is; that is the precondition the framework did not expose. Evidence about a run and a
reference state to test the oracle against are different requirements, and our negative result is
about the second.

All the remaining results come from one agent benchmark with two task domains, four agents and six
simulator models, all served by one provider. The trajectory-level effect is carried by one agent; the others
have a zero base rate in both arms and cannot show it, so the mechanism's generality across models
is inferred from its mechanics rather than observed. The second domain was intended to relax this
and did not: the cross-vendor agent collapsed in none of its pilot episodes, so the arm was dropped
and \emph{independent-model validation remains open}. The cross-lingual setting is a simulator
speaking Chinese against an English tool environment; other language pairs are untested, and the
simulated users are language models rather than human speakers, which leaves open how much of the
effect belongs to the simulator's language behaviour and how much to the agent's. Inference is
conditional on the sampled tasks: the retail results are conditional on the 30 tasks drawn from the
benchmark's test suite, and the airline results on all 50 of that domain's tasks.

\subsection{Conclusion validity}
The primary inference is a task-clustered interval, and the reference distribution is $t(G-1)$ with
$G$ the number of tasks, which treats tasks as the sampling unit. This is the right unit for claims
about a task suite and the wrong unit for claims about episodes; the intervals we report are
conditional on the tasks sampled and do not absorb the sampling step. Exact tests are reported only
as a cross-check and are not used to declare an effect resolved. Several of our secondary analyses
--- permutation, sign-flip and task-clustered bootstrap variants --- do not share a single null
hypothesis and therefore do not constitute independent lines of evidence; they are reported as
sensitivity analyses of one estimate. Finally, an interval that contains zero is reported as no
detected difference and not as evidence of no difference.

\section{Conclusion}\label{sec:conclusion}
An agent evaluation harness is a piece of software, and the score it produces is an instrument
reading. We audited one such instrument and report two findings. Its test oracle mutated the state
it was grading, which inflated the recorded success rate from 65.5\% to 94.3\% in our runs and would
have done so silently; the mechanism is a missing restore, the detection is a state-comparing
repeat-scoring check or a terminal-state check, and the correction is a replay of the recorded
actions rather than a re-run.
The correction is also not a uniform offset: it removes between 24 and 34 percentage points from
each agent measured here and reorders the top of the matrix: the inflated score and the corrupted
comparison are two consequences of the same failure. A pre-registered ablation attributes 20
additional target--variant detections to exception restoration, order dependence and format
tolerance at about 1.7--1.8$\times$ the basic-check cost, while an out-of-design over-lenient numeric
matcher is detected by no check. And the language of the apparatus is a configuration
parameter of the measurement rather than a property of the model: localising the agent's own world
is associated with a change in the rate at which trajectories degenerate, each of the three
documents the agent reads is associated with a lower rate when localised alone in the tested
configuration, and the effects do not combine additively. The relative size of
the three is an observed ordering in this configuration, not an established mechanism. The sign of
that effect does not survive a change of task domain. A
score table that reports neither the trajectory-collapse rate nor the configuration it was taken
under is reporting less than it appears to.

\backmatter

\bmhead{Declarations}
\begin{itemize}
  \item \textbf{Funding.} No funding was received for conducting this study.
  \item \textbf{Competing interests.} The author declares no competing interests.
  \item \textbf{Ethics approval.} This study involves no human participants and no animal
        subjects. The ``users'' in all experiments are language-model simulations; no human
        subjects were recruited, tested, or recorded.
  \item \textbf{Data availability.} The software artifact and the frozen evaluation dataset are
        archived separately on Zenodo. The software artifact is at
        \url{https://doi.org/10.5281/zenodo.23290600} (v1.0.0). It contains the oracle-integrity
        suite, the fault-injection battery, the analyses that produce the reported tables and
        data-derived figures in this paper, the version-binding manifests, and the small frozen
        inputs those analyses read.

        Behind those analyses are four frozen archives: the v1.0 main experiment (4{,}320 episodes)
        with both the archived and the state-corrected labels; the S2 release holding the
        component ablation (3{,}240 cells), the cross-domain experiment (1{,}200 cells) and the
        blind-adjudication inputs (212 items); the enhancement round holding the held-out variant
        battery, the detector ablations, the power analysis and the cross-framework portability
        attempt; and an anonymised copy of the first. They total 232 MB and are archived separately
        at \url{https://doi.org/10.5281/zenodo.23290703}, together with the data terms they are
        released under and a checksum file. Each number in this paper is bound to exactly one of
        those archive revisions, and the mapping from archive to experiment to section, with
        SHA-256 checksums, is recorded in the package's release manifest.

        Two analyses read only the archived and state-corrected labels of the v1.0 archive: the
        transition matrix between the two labels and the comparison of agents under each
        (Section~\ref{sec:res-oracle}). Their scripts, outputs and source hashes are included with
        this preprint as the ancillary file \texttt{simglot-v2-analysis.zip}: a README with the two
        commands, a source-binding file recording the SHA-256 of the v1.0 archive and of the replay
        table the scripts read, the four output files, and a manifest. Each analysis is bound to the
        frozen evidence it reads.

        The pre-registered follow-up ablation (P0 in the project files) is attached as
        \texttt{simglot-p0-archive-20261125.tar.gz} (80{,}431 bytes; SHA-256
        \texttt{9fb6f468c16861e8}\allowbreak\texttt{09d0bee5754489b7}\allowbreak\texttt{95f0902ad5f0e3d}\allowbreak\texttt{d1248619a8e06a05f}).
        It contains the frozen protocol (SHA-256
        \texttt{8dfb69cf81ffe5f6}\allowbreak\texttt{6325e9d5ec353896}\allowbreak\texttt{773f9a625240fcd5}\allowbreak\texttt{37776ae7f11706e6}),
        isolated driver and variant builder, hash-pinned upstream sources and adaptation diff,
        full and smoke outputs, run records, and \texttt{MANIFEST.sha256}. Its 21 payload files plus
        the manifest (22 files total) all verify. The run made no model calls and isolated each
        method $\times$ variant $\times$ target cell in a subprocess.

        Two things are deliberately not distributed. The second, independently implemented agent
        framework used for the portability result ships its task data under a licence that permits
        redistribution only in encrypted form, so the package carries the adapter, the applicability
        findings and the version fingerprint, and this paper reports no detection rate for it. The
        model-served episode records are distributed as frozen records; they contain outputs of
        third-party model services and are not offered under an unconditional open licence.

        The code authored for this study is released under the \textbf{MIT} licence. The frozen
        records are \emph{not} covered by that licence and carry their own terms, stated in the
        package: the derived results and the per-episode records may be used and redistributed for
        research with attribution, while large-scale redistribution or commercial use of the
        episode records should be requested first, because they contain outputs of third-party
        model services.

        Note on byte-reproducibility: the analyses are deterministic from the frozen records, but
        the \emph{PDF} is not byte-reproducible because the \LaTeX{} engine writes a build timestamp
        and document identifier. The published PDF is bound to the released source by a source hash,
        not a PDF hash.
  \item \textbf{Code availability.} The code is in the archives above and at
        \url{https://github.com/JunjieChen0/SimGlot} (tag \texttt{v1.0.0}). Three offline entry points
        cover the S2, enhancement and v1.0 analyses; the follow-up ablation archive has its own. The
        oracle-integrity suite runs in one command without network, credentials or a model, and prints
        the matrix in Table~\ref{tab:kit}. The end-to-end analysis runs from archived episodes and
        refuses to run if an API key is present. Its four-operation interface can be pointed at another
        harness's oracle without changing the suite.
  \item \textbf{Use of large language models.} Language models are the experimental material of
        this study: they act as the tool-calling agents under test, as the user simulators that
        generate the test inputs, as the translator that produced the Chinese task instructions
        (with the mechanical fidelity check described in Section~\ref{sec:bench}), and as the
        independent adjudicator in Section~\ref{sec:res-general}. The models are specified by name
        and version throughout, and the prompts given to the adjudicator and to the translator are
        released with the package. Language models were also used to assist in drafting the
        manuscript and the analysis code; the author reviewed and is responsible for all content,
        analyses and conclusions.
\end{itemize}

\bibliography{refs}

\end{document}